\documentclass[12pt]{article}
\usepackage{graphicx}
\usepackage{epsfig,a4}
\begin{document}
\title{Diffractive processes at high energies. \footnote{To be published in proceedings of Quarks-2004}}
\author{Abramovsky V.A.\footnote{ava@novsu.ac.ru}, Dmitriev A.V.\footnote{gridlab@novsu.ac.ru}, Shneider A.A. \\ \footnotesize Novgorod State University, B. S.-Peterburgskaya Street 41,\\
\footnotesize Novgorod the Great, Russia, 173003}
\maketitle
\begin{abstract}
In this work we calculate pomeron flux in the single diffraction processes. We consider two models: quasi-eikonal model and low constituent model. Both models give the pictures different from the traditional  three-reggeon model. Successive developing of modeles gives some indications, that the low constituent model is more attractive.
\end{abstract}

\section{Introduction}
\begin{figure}
\includegraphics[scale=0.35]{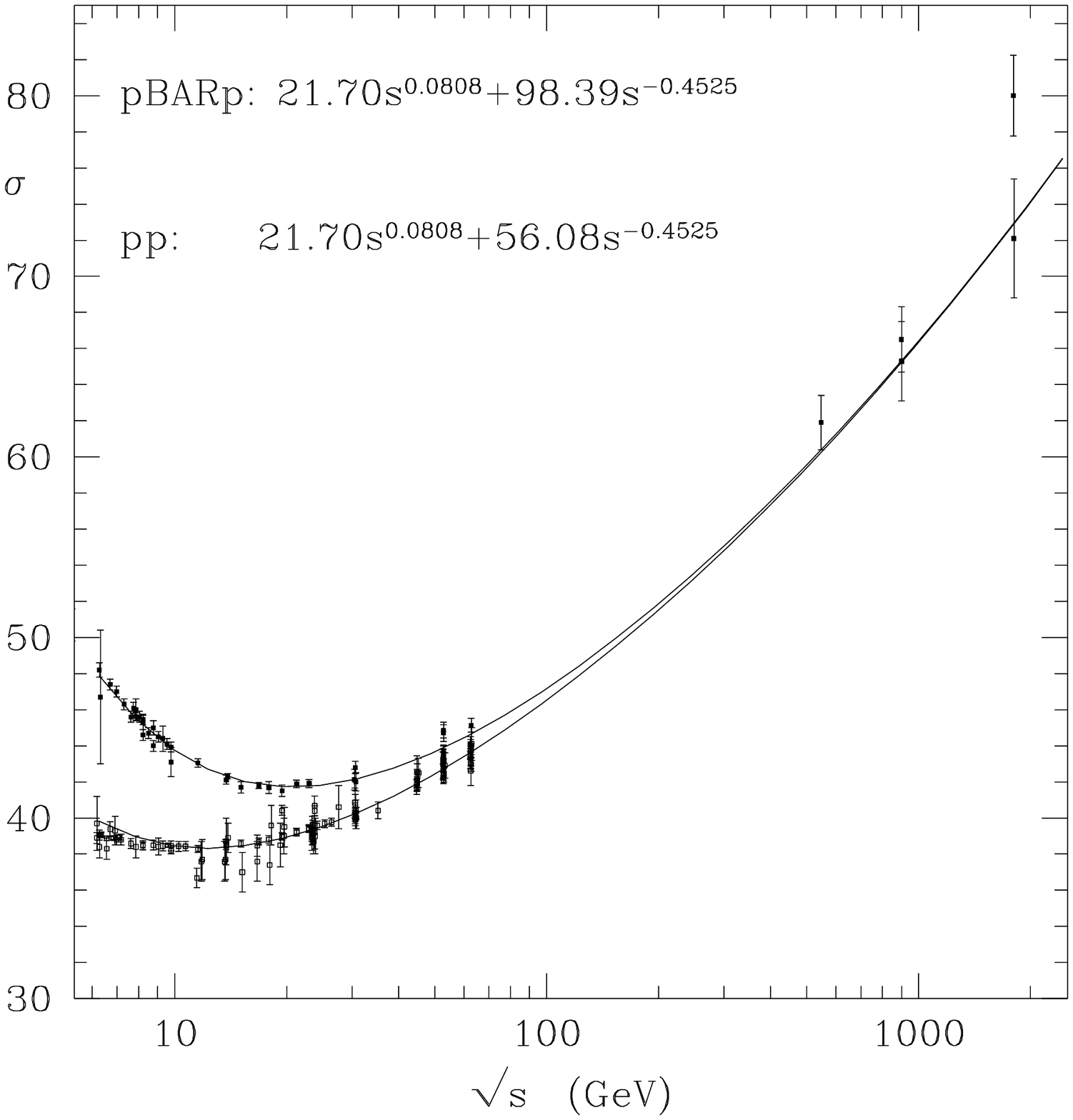}
\includegraphics[scale=0.35]{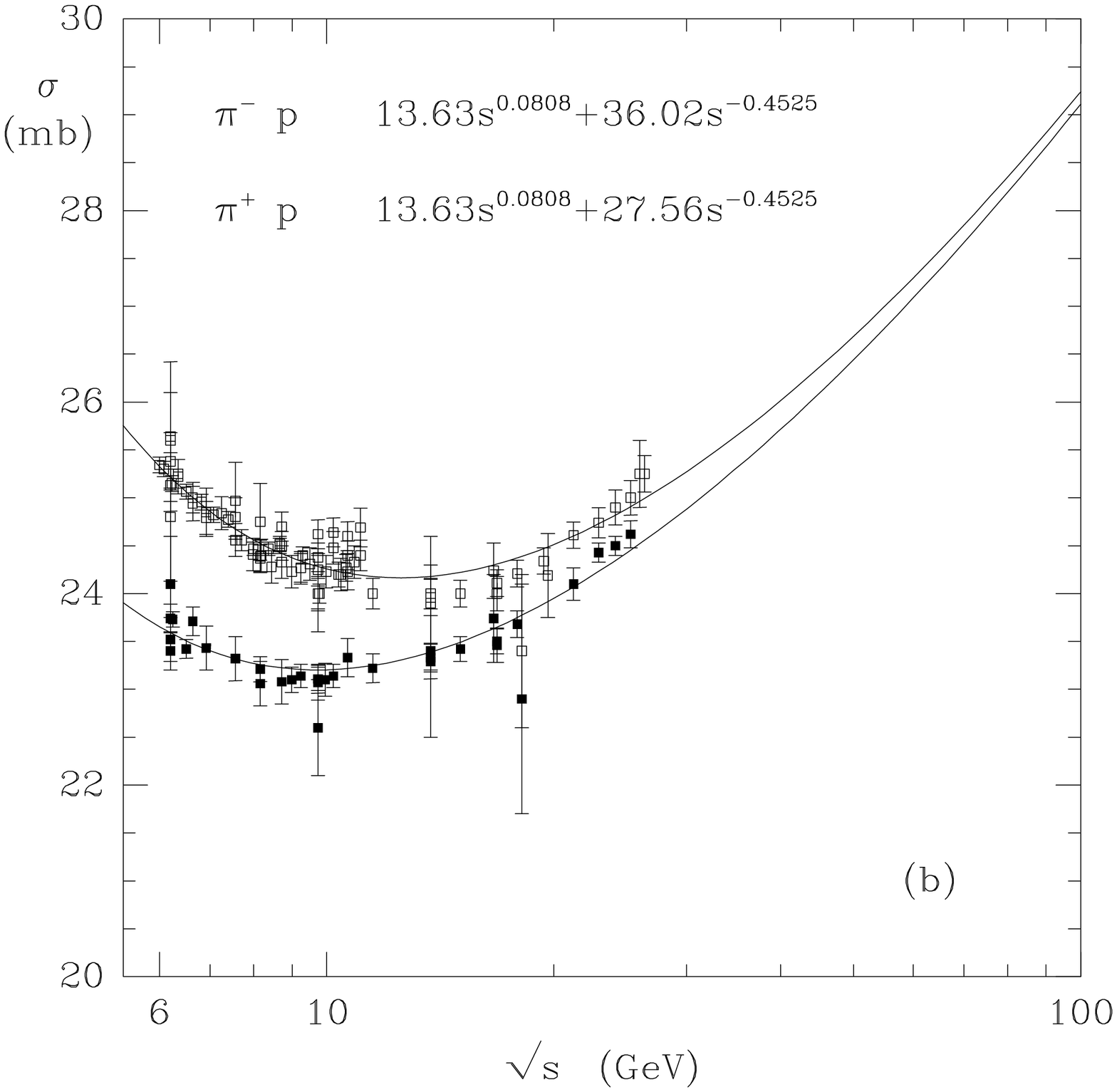}
\label{fig:sigtot}
\caption{Total cross-sections in the Donnachie-Landshoff parametrization.}
\end{figure}

\begin{figure}
\begin{center}                                                               
\input{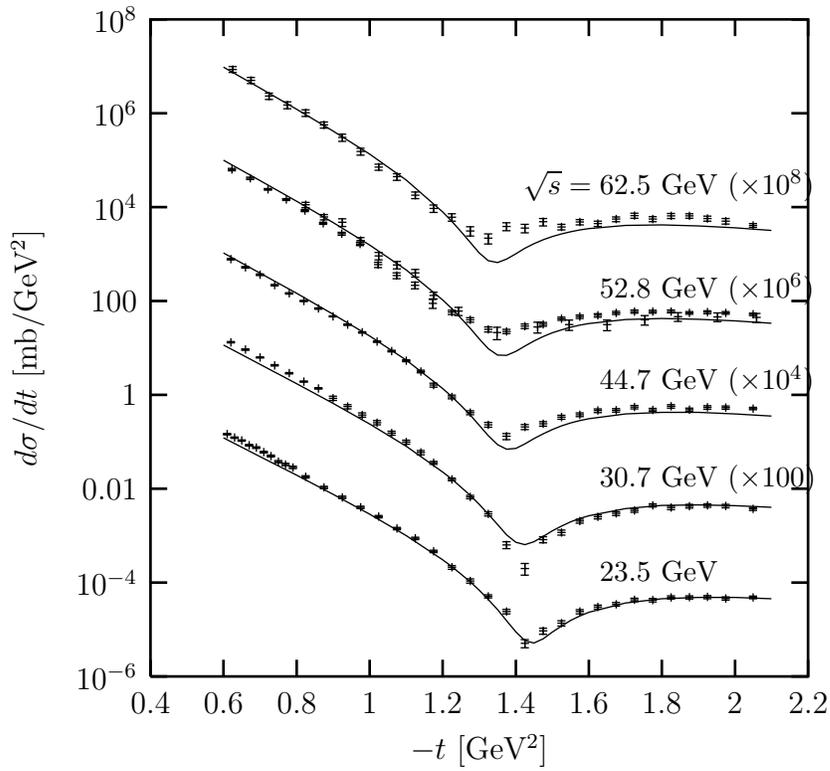}
\end{center}
\caption{The Donnachie-Landshoff fit for the differential elastic $pp$ cross section}
\label{fig:sigel}
\end{figure}

Regge non-enhanced phenomenology well describes total and elastic cross-sections in the Donnachie-Landshoff parametrization \cite{DL}, see Fig.\ref{fig:sigtot}, taken from \cite{DLfig} and Fig.\ref{fig:sigel}, taken from  \cite{elfig}.

\begin{figure}
\begin{center}
\includegraphics[scale=0.5]{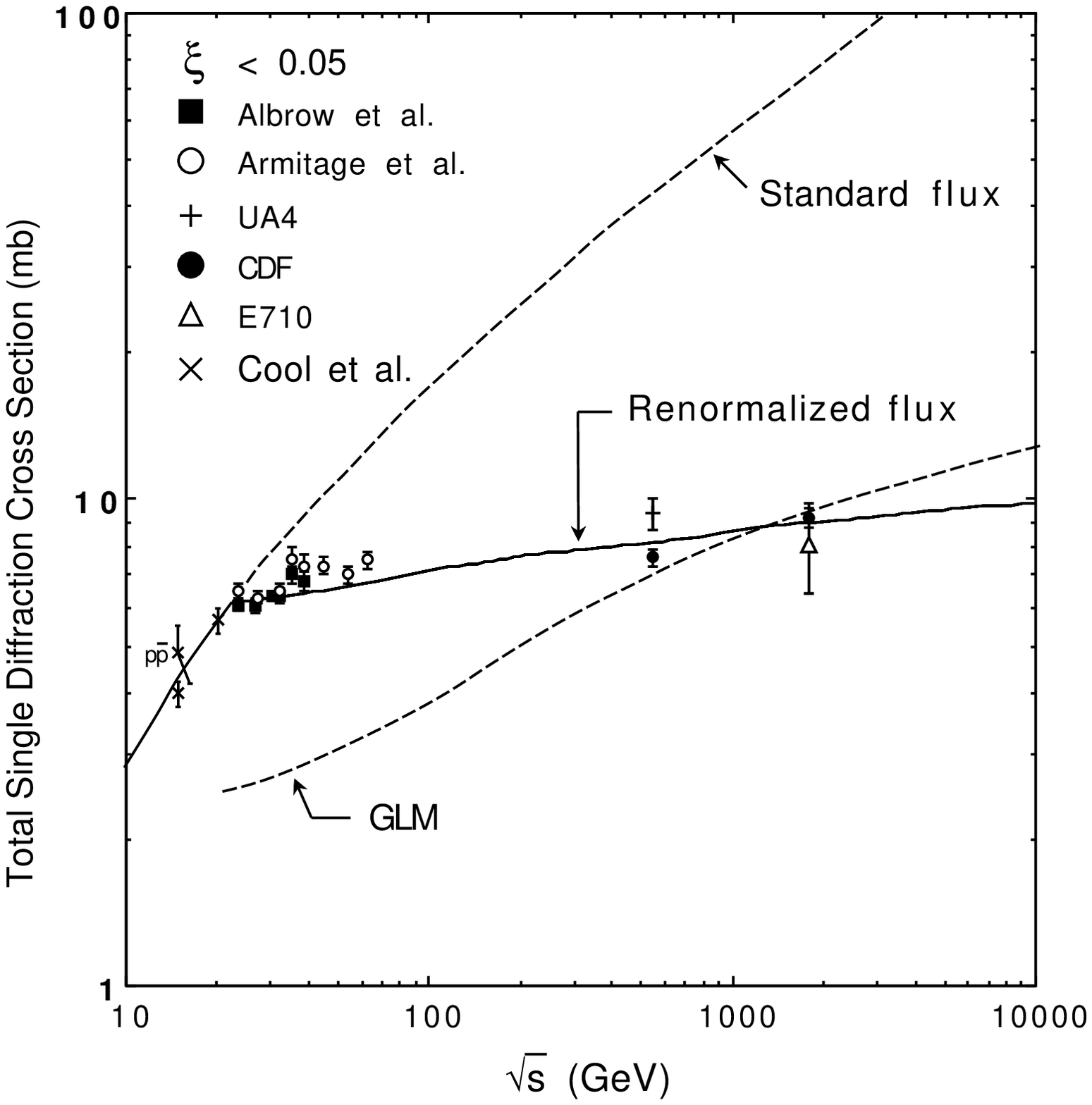}
\vspace*{0pt}
\caption{The total single diffraction cross section for $p(\bar p)+p\rightarrow p(\bar p)+X$ vs $\sqrt s$ compared with the predictions of the renormalized pomeron flux model of Goulianos~\cite{goul_orig} (solid line) and the model of Gotsman, Levin and Maor~\cite{gotsman_orig} (dashed line, labeled GLM); the latter, which includes "screening corrections", is normalized to the average value of the two CDF measurements at $\sqrt s=546$ and 1800 GeV.}
\label{fig:figgoul}
\end{center}
\end{figure}

Low-energy single diffraction data is also well described by regge phenomenology with supercritical pomeron, but at the region of Tevatron
energies it fails to describe data on single diffraction dissociation. The main problem is that total single diffraction
cross-section rise considerably weaker than it is predicted by $Y$-like Regge diagrams involving only three pomerons. This fact is
clearly seen from Fig.\ref{fig:figgoul} extracted from Ref.\cite{goul_pic}, there "Standard flux" corresponds to $Y$-like Regge diagram.

Many ways were suggested to solve this problem. First way is two-variant (Ref.\cite{goul_orig} and Ref.\cite{erhan_orig}) pomeron flux renormalization model, where we consider the equation for cross section of single diffraction
\begin{equation}
\frac{d^3\sigma}{dM^2dt}=f_{I\!\!P /p}(x,t)\sigma_{I\!\!Pp}(s)
\end{equation}
and pick out the factor, named as 'pomeron flux'
\begin{equation}
f_{I\!\!P /p}(x,t)=K\xi^{1-2\alpha_{I\!\!P}(t))}
\end{equation}
Renormalization of the pomeron flux is made by intserting dependence of $K$ either on $s$ (as in Ref.\cite{goul_orig}) either on $x$,$t$, as  in Ref.\cite{erhan_orig}.
This phenomenological approach well describes CDF data on single diffraction, but we need more theoretical bases for extrapolation to higher energies.

The second way is straight-forward account of screening corrections (Ref.\cite{gotsman_orig} and Ref.\cite{chung_orig}).
This way seems to be more natural, but we need to introduce additional parameters and make some assumptions about Regge
diagram technics. In Ref.\cite{gotsman_orig} and Ref.\cite{chung_orig} only some parts of sufficient diagrams  were
accounted by going to the impact parameter space $b$ and replacement initial "Borhn" factor $\chi(s,\overline{b})$ to eikonalized amplitude $(1-e^{-\mu\chi(s,\overline{b})})$. In addiction, the central $Y$-like diagram was modified to account low-energy processes and in Ref.\cite{chung_orig} the dependence of pomeron intercept on energy was introduced. 

As compared  with Ref.\cite{gotsman_orig} and Ref.\cite{chung_orig} we successivly consider all non-enhanced diagrams. 

In this work we also consider low constituent model, where there is only basic quark-gluon states and interactions. This model leads us to the non-local pomeron, but it has clear interpretation of the pomeron flux.

\section{Quasi-eikonal model}

Quasi-eikonal model, considered in this work, is standart enough. We use reggeon diagram technic with reggeon propogator $s^{\alpha(t)}$, model gauss vertexes of the interaction of n pomerons with hadron 
\begin{equation}
N_h(k_1,..,k_n)=g_h(g_hc_h)^{n-1}exp\left(-R_h^2\sum_{i=1}^{n}k_i^2\right)
\end{equation}
and the vertex corresponding to the transition of $l$ pomerons into $m$ pomerons under the $\pi$ -meson exchange dominance assumption
\begin{equation}
\Lambda(k_1,..,k_{m})=r(g_{\pi}c_{\pi})^{m-3}exp\left(-R_r^2\sum_{i=1}^{m}k_i^2\right).
\end{equation}
Here $g_h$ is the pomeron-hadron coupling, $c_h$ is the corresponding shower enhancement coefficient, $R_h$ and $R_r$ are
the radii of the pomeron-hadron and pomeron-pomeron interactions, respectively, $k_i$ are the pomeron transverse momenta.
Integration on the pomeron momenta is made trivial in the impact parameter space represenztation and we only have to sum on then nubmers of pomerons, attached to the same vertexes.

\begin{figure}
\begin{center}
\includegraphics[scale=0.5,angle=90]{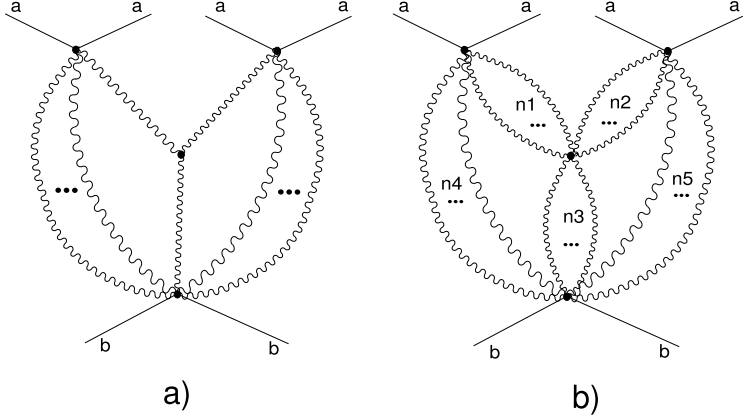}
\vspace*{0pt}
\caption{Regge diagrams describing single diffraction dissociation of particle b.}
\label{fig:fig2}
\end{center}
\end{figure}

As compared with Ref.\cite{gotsman_orig} and Ref.\cite{chung_orig}, where only part of sufficient diagrams  was accounted ( see Fig. \ref{fig:fig2}a), in this paper we account all non-enhanced absorptive corrections to the Y-diagram
contribution, shown in  Fig.\ref{fig:fig2}b.

Because low-energy corrections rapidly decrease with energy, we account only pomeron contributions, but in all
sufficient diagrams, as it was done in Ref.\cite{all} and Ref.\cite{Ab}. It gives us possibility to normalize cross-section of the single diffraction to CDF data and make theoreticaly based predictions for cross-section of the single diffraction at LHC energies.

The contribution $f_{n_1 n_2 n_3 n_4 n_5}$ of each diagram in Fig.\ref{fig:fig2}b can be written in a \begin{it} rather simple form \end{it}
 \begin{equation}
\begin{array}{l}
f_{n_1 n_2 n_3 n_4 n_5}=\frac{(-1)^{n_1+n_2+n_3+n_4+n_5+1}}{n_1!n_2!n_3!n_4!n_5!}\frac{8 \pi^3 r}{c_a^2 c_b g_{\pi} c_{\pi}}
\left[\frac{g_a c_a g_{\pi} c_{\pi} e^{\Delta (Y-y)}}
{8\pi (R_a^2+R_{\pi}^2+\alpha^{\prime}(Y-y))}\right]^{n1+n2} \\
\\
\left[\frac{g_a c_a g_b c_b e^{\Delta Y}}
{8\pi (R_a^2+R_b^2+\alpha^{\prime}Y)}\right]^{n4+n5}
\left[\frac{g_b c_b g_{\pi} c_{\pi} e^{\Delta y}}
{8\pi (R_b^2+R_{\pi}^2+\alpha^{\prime}y)}\right]^{n3}
\frac{1}{detF}e^{-t\frac{c}{detF}} \\
\\
detF=a_1 a_2 a_3 + a_1 a_3 a_5 + a_1 a_2 a_5 + a_1 a_2 a_4 + a_2 a_3 a_4 + a_1 a_4 a_5 + a_3 a_4 a_5 + a_2 a_4 a_5 \\
c=a_2 a_3 + a_1 a_5 + a_3 a_5 + a_2 a_5 + a_1 a_3 + a_1 a_4 + a_3 a_4 + a_2 a_4 \\
a_1=\frac{n_1}{R_a^2+R_{\pi}^2+\alpha^{\prime}(Y-y)}\\
a_2=\frac{n_2}{R_a^2+R_{\pi}^2+\alpha^{\prime}(Y-y)}\\
a_3=\frac{n_3}{R_b^2+R_{\pi}^2+\alpha^{\prime}y}\\
a_4=\frac{n_4}{R_a^2+R_b^2+\alpha^{\prime}Y}\\
a_5=\frac{n_5}{R_a^2+R_b^2+\alpha^{\prime}Y}.
\end{array}
\end{equation}
Here $Y=ln(s)$ $y=ln(M^2)$.
Then inclusive cross section is
\begin{equation}
(2\pi)^3 2E \frac{d^3\sigma }{dp^3 }=\pi \frac{s}{M^2}\sum_{n_1,n_2,n_3=1}^{\infty} \sum_{n_4,n_5=0}^{\infty}f_{n_1 n_2 n_3 n_4 n_5}
\end{equation}

Our method differs from early work Ref. \cite{Ab}, where all parameters but vertex $r$ were fixed. In this work here we vary all parameters. Parameters were varied with natural limitations, i.e. all parameters were varied above its conventional values. We don`t consider very high or very low values of pomeron itercept and slope, which can be compensated by other parameters.

Another difference as compared with Ref.\cite{Ab} is the fact, that we use data on total and elestic (differential) cross-sections and data on total single-diffraction cross-sections. So, we can fix parameters of the model with higher precision and with account of its one-to-one corellations.

The model under consideration doesn`t include possible contributions of low-lying reggeons, so we limit considered energies by $\sqrt{s}>52 GeV$ for elastic and total cross-sections. As was shown in \cite{ejela}, modern data don`t give us possibility to distinct simple ploe model with total cross-sections  $\sigma_{tot}=As^\Delta$ and eikonaliezed models with  $\sigma_{tot}=C+Dln(s)$ or $\sigma_{tot}=E+Fln(s)^2$. But we can reliably determine parameters of the model $R_h,g_h,\Delta,\alpha^{\prime}$ from elastic and total cross-section data at fixing $c_h$.

We use CDF data on single diffraction for analysis.
 
CDF data \cite{CDF} was presented as a result of the monte-carlo simulations based on the general formula:
\begin{equation}
\begin{array}{l}
\frac{d^2\sigma}{d\xi dt}
=\frac{1}{2}\left[\frac{D}{\xi^{1+\epsilon}}
e^{\textstyle (b_0-2\alpha'_{SD}\ln \xi)t} 
+I\xi^{\textstyle \gamma}e^{\textstyle b't}\right]
\label{CDFfit} \\
\xi \equiv 1-x
\end{array}
\end{equation}

Taken CDF data is shown in Table 1.

\begin{table}[h]
\caption{CDF fit-parameters from reference~[1].}
\begin{center}
\begin{tabular}{|c|c|c|}
\hline
         &                       &                        \\
         & $\sqrt{s}=546\,\,GeV$ & $\sqrt{s}=1800\,\,GeV$ \\
         &                       &                        \\ \hline
         &                       &                        \\
$D$      & $3.53 \pm 0.35$       & $2.54 \pm 0.43$        \\
$b_{0}$  & $7.7 \pm 0.6$          & $4.2 \pm 0.5$          \\
$\alpha'_{SD}$ & $0.25 \pm 0.02$      & $0.25 \pm 0.02$        \\
$\epsilon$ & $0.121 \pm 0.011$   & $0.103 \pm 0.017$      \\
$I$      & $537^{+498}_{-280}$    & $162^{+160}_{-85}$       \\
$\gamma$ & $0.71 \pm 0.22$       & $0.1 \pm 0.16$         \\
$b'$     & $10.2 \pm 1.5$        & $7.3 \pm 1.0$          \\
         &                       &                        \\ \hline
\end{tabular}
\end{center}
\end{table}

This parameters are experimental points tested in our model. Let`s mark, that low-lying reggeons contribution, corresponding to second addendum in  (\ref{CDFfit}), isn`t acconted in our calculations and we have to model only parameters  $D$,$b_{0}$,$\alpha'_{SD}$,$\epsilon$. We calculate this parameters in the region ${0.05<t<0.1; 0.99<x<0.995}$, where we have the most reliable CDF data and contribution of the low-lying reggeons is mnimal.

We have to mark, that this data is not precise because of the following reasons:
 \begin{enumerate}
 \item CDF single diffraction data have low statistcs and narrow kinematical region, where the data was taken;
 \item At each energy ($\sqrt{s}=546GeV$ É $\sqrt{s}=1800GeV$) 6 highly correlated parameters are introduced, and it makes calculations unstable;
 \item Fixing of effective pomeron slope on the common value  $\alpha'_{SD}=0.25$ is obliged.
\end{enumerate}
  
Unreability of the data in Table1 is cearly seen from analysis of dependence of  $D$ on energy from $\sqrt{s}=546GeV$ to $\sqrt{s}=1800GeV$. As defined \cite{CDF}, 
\begin{equation}
D=G(0)s^{\Delta}
\end{equation}
here G(0) doesn`t depend on $s$, and $\Delta>0$. In accordance with this definition, parameter $D$ must increase when energy increases, but in  CDF data it decreases.

Total single diffraction cross-sections are well experimentally defined and don`t depend on the model, used in analysis of basic data (detectors counts)
\begin{equation}
\begin{array}{ll}
\sigma_{SD}(\sqrt{s}=546 GeV)=7.89 \pm 0.33 mb \\
\sigma_{SD}(\sqrt{s}=1800 GeV)=9.46 \pm 0.44 mb
\end{array}
\label{sig_SD_exp}
\end{equation}

We include these two points in  $\chi^2$ test, but with larger weights, than points shown in Table 1.

Because total and elastic cross sections, on one side, and single diffraction cross sections, on other side, have different types, we vary parameters  $r$,$R_{\pi}$ and $c_{\pi}$ to achieve the best agreement with data in Table1 and data (\ref{sig_SD_exp}), fixing at each step parameters  $\Delta$,$\alpha^{\prime}$,$g_h$ and $R_h$ from total and elastic cross-sections.

Results.

In the end of optimization process we`ve got next parameter set:
$g_p^2=  75.0538,  \Delta=  0.0868089,  R_p^2=1.94755, \alpha\prime=0.148963, 
c_p^2=2.03954,  r=0.111525,  R_{\pi}^2=0.173682,  c_{\pi}^2=6.00989$

Following total single diffraction cross sections were calculated at these parameters:

\begin{equation}
\begin{array}{ll}
\sigma_{SD}(\sqrt{s}=546 GeV)=7.5 mb \\
\sigma_{SD}(\sqrt{s}=1800 GeV)=10 mb
\end{array}
\label{sig_SD_model}
\end{equation}

Corresponding differential characteristics of differential single diffraction cross sections are enumerated in Table 2:

\begin{table}[h]
\caption{Differential characteristics of differential single diffraction cross sections in our model.}
\begin{center}
\begin{tabular}{|c|c|c|}
\hline
         &                       &                        \\
         & $\sqrt{s}=546\,\,GeV$ & $\sqrt{s}=1800\,\,GeV$ \\
         &                       &                        \\ \hline
         &                       &                        \\
$D$      & $2.9628$       & $3.08731$        \\
$b_{0}$  & $5.32553$          & $5.27187$          \\
$\alpha'_{SD}$ & $0.294999$      & $0.270871$        \\
$\epsilon$ & $0.0580572$   & $0.0549202$      \\
         &                       &                        \\ \hline
\end{tabular}
\end{center}
\end{table}

Calculated differential characteristics are very close to ones from the triple-pomerom model, so the following relation is satisfied
\begin{equation}
\frac{d^3\sigma}{dM^2dt}=f_{I\!\!P /p}(x,t)\sigma_{I\!\!Pp}(s)
\end{equation}
where 
\begin{equation}
f_{I\!\!P /p}(x,t)=K(s)\xi^{1-2\alpha_{I\!\!P}(t))}
\end{equation}
is renormalized pomeron flux. As compared with standart triple-pomeron model the dependence of factor $K$ on ernergy $s$ is introduced. This dependence provides slowing on the rise of the single diffraction cross section with energy.

\begin{figure}
\begin{center}
\includegraphics[scale=0.75]{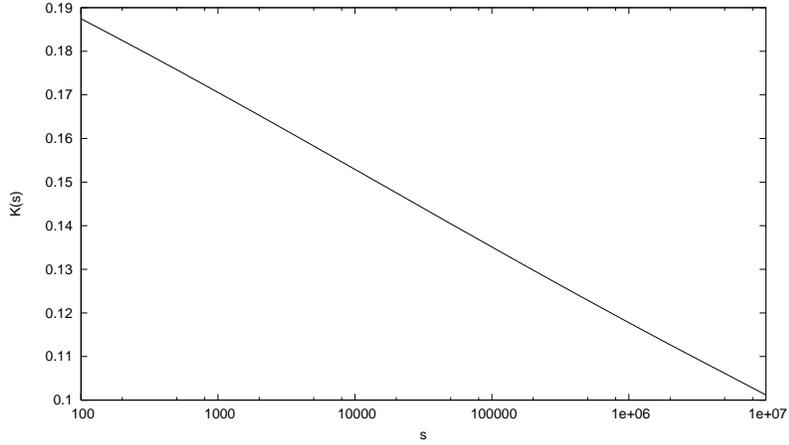}
\vspace*{0pt}
\caption{Dependence of renormalizing factor $K(s)$ on energy.}
\label{fig:K_s}
\end{center}
\end{figure}

Dependence of renormalizing factor $K(s)$ on energy is shown on Fig.\ref{fig:K_s}.

\begin{figure}
\begin{center}
\includegraphics[scale=0.75]{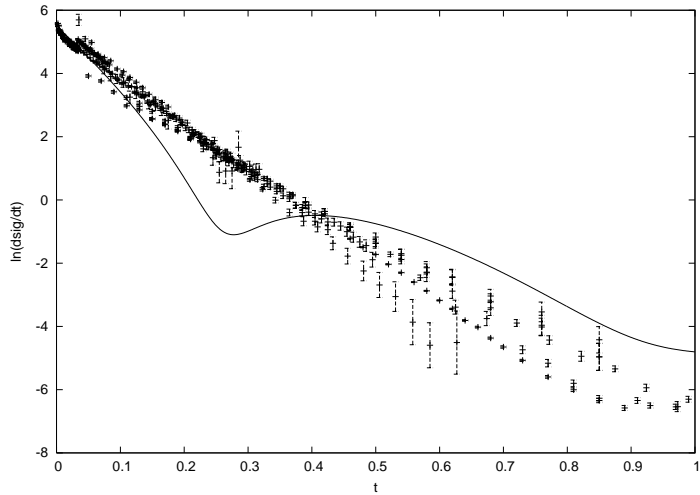}
\vspace*{0pt}
\caption{Elastic cross sections $\frac{d\sigma}{dt}$ for reaction $p+p \rightarrow p+p$. Theoretical curve is at energy $\sqrt{s}=1800GeV$. Experimental points are taken at energies from ISR to Tevatron.}
\label{fig:elastic}
\end{center}
\end{figure}

We have to note, that we have inconsistences that calculating $c_p$. From one side, there are theoretical indications, that $c_p>1$. Such high values of $c_p$  lead to significant divergence of dependence $\frac{d\sigma}{dt}$ on $t$ from exponential behavior $e^{-bt}$ already at  $t{\sim}0.2 GeV^2$. It is known from experiment, that elastic cross-section falls exponentially on $t$ up to $t{\sim}1GeV^2$. This inconsistence is clearly seen from Fig.\ref{fig:elastic}.

From this fact of independence of logarithmic slope on $t$ we conclude, that $c_p<<1$. To explain slow rise of  $\sigma_{SD}$ with energy we have to assume very high $c_{\pi}$, $c_{\pi}c_p \gg 1$ at $g_{\pi} \sim g_{p}$. It gives desired value of the fraction  $\frac{\sigma(\sqrt{s}=1800GeV)}{\sigma(\sqrt{s}=546GeV)} \sim 1.2$, but leads to very high values of logarithmic slope  $b \sim 50 GeV^{-2}$ (situation will be even more worse, than in the case of elastic cross-section, shown on Fig.\ref{fig:elastic}). Solving of this problem by precise adaptation of $R_{\pi}$ is unusable, because it leads to highly differing from experiment and depended on $M^2$ and $t$ values of  $\alpha^{\prime}$ and $\epsilon$.

At $c_p>1$ we don`t need $c_\pi$ in so high values, and logarithmic slope $b$ is back to values about ones, not tens. So, we must return to the theoretically based area  $c_p>1$ and limit considered area of elastic scattering by $t<0.2 GeV^2$.

\begin{figure}
\begin{center}
\includegraphics[scale=0.8]{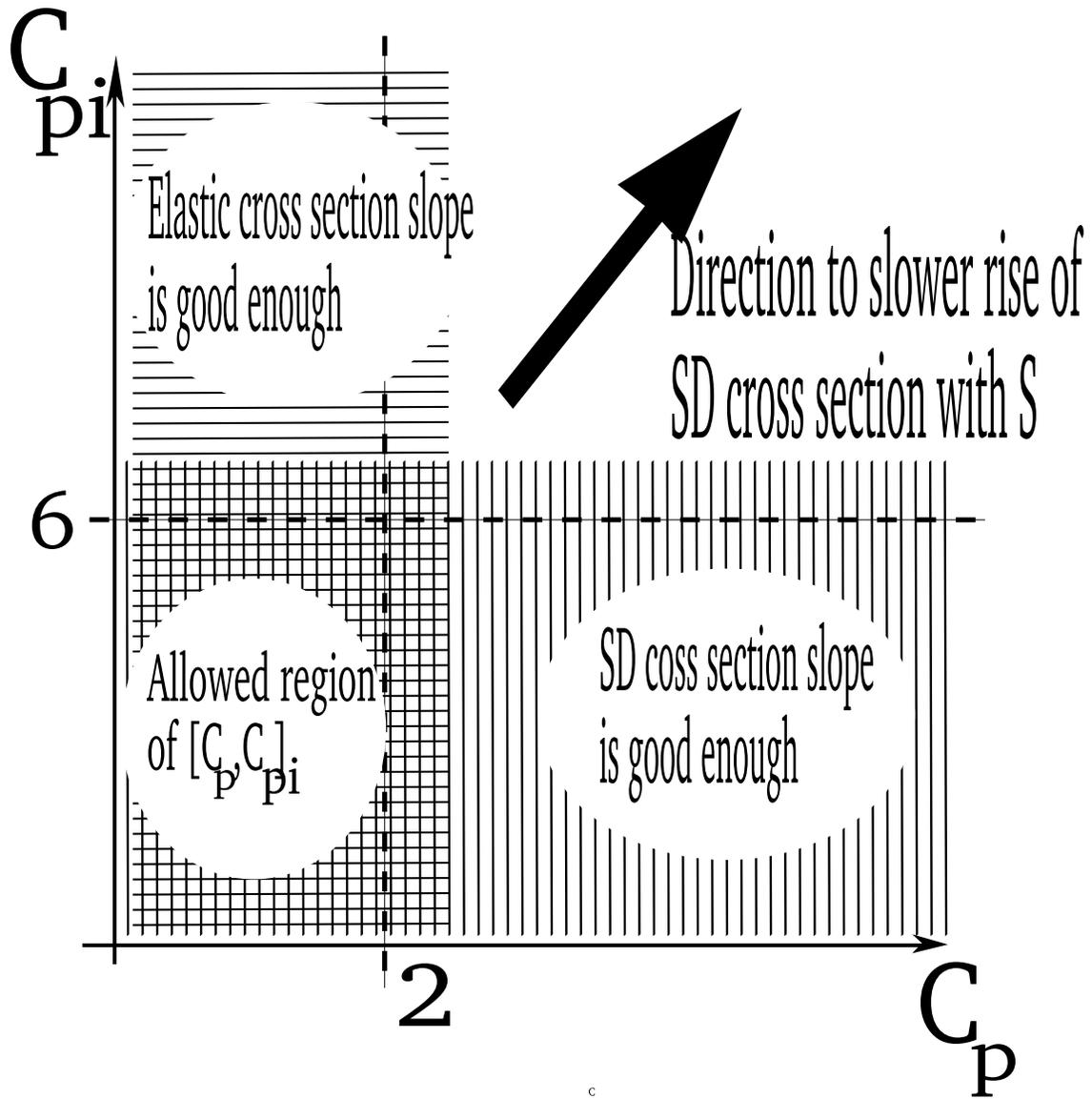}
\vspace*{0pt}
\caption{Allowed region of parameters $[c_p,c_\pi]$.}
\label{fig:call}
\end{center}
\end{figure}

We see, that goog agreement of quasi-eikonal model with experiment is achieved on the border of the allowed region of parameters  $[c_p,c_\pi]$ (see. Fig.\ref{fig:call}).

 From one side, it gives stability of the calculated parameters. From the other side this model has no reserve of stability. If fraction of the cross sections $\frac{\sigma(\sqrt{s}=1800GeV)}{\sigma(\sqrt{s}=546GeV)}$ will be defined more precisely and will be found in the region $1.1 \div 1.15$ (it is minimal value, which is consistent with existed data), then for description of this data we will be obliged to decline either describing elastic and total cross sections or describing logarithmic slope of the single diffraction on $t$.
 
\newpage
\section{Low constituent model}
We consider the three-stage model of hadron interaction at the high energies.

On the first stage before the collision there is a small number of partons in 
hadrons. Their number, basically, coinsides with number of  
valent quarks and slow increases with the rise of energy owing to the appearance 
of the  breasstralung gluons.

On the second stage the hadron interaction is carried out by gluon exchange between the 
valent quarks and initial (bremsstralung) gluons and the hadrons gain the colour 
charge. 

On the third step after the interaction the colour hadrons fly away and when the 
distance between them becomes more than the confinement radius $r_c$, the lines 
of the colour electric field gather into the tube of the radius $r_c$. This tube 
breaks out into the secondary hadrons.

Because the process of the secondary harons production from colour tube goes 
with the probability 1, module square of the inelastic amplitudes corresponds 
to the elastic amplitude with the different number partons.  

\begin{figure}
\includegraphics[width=2.5in]{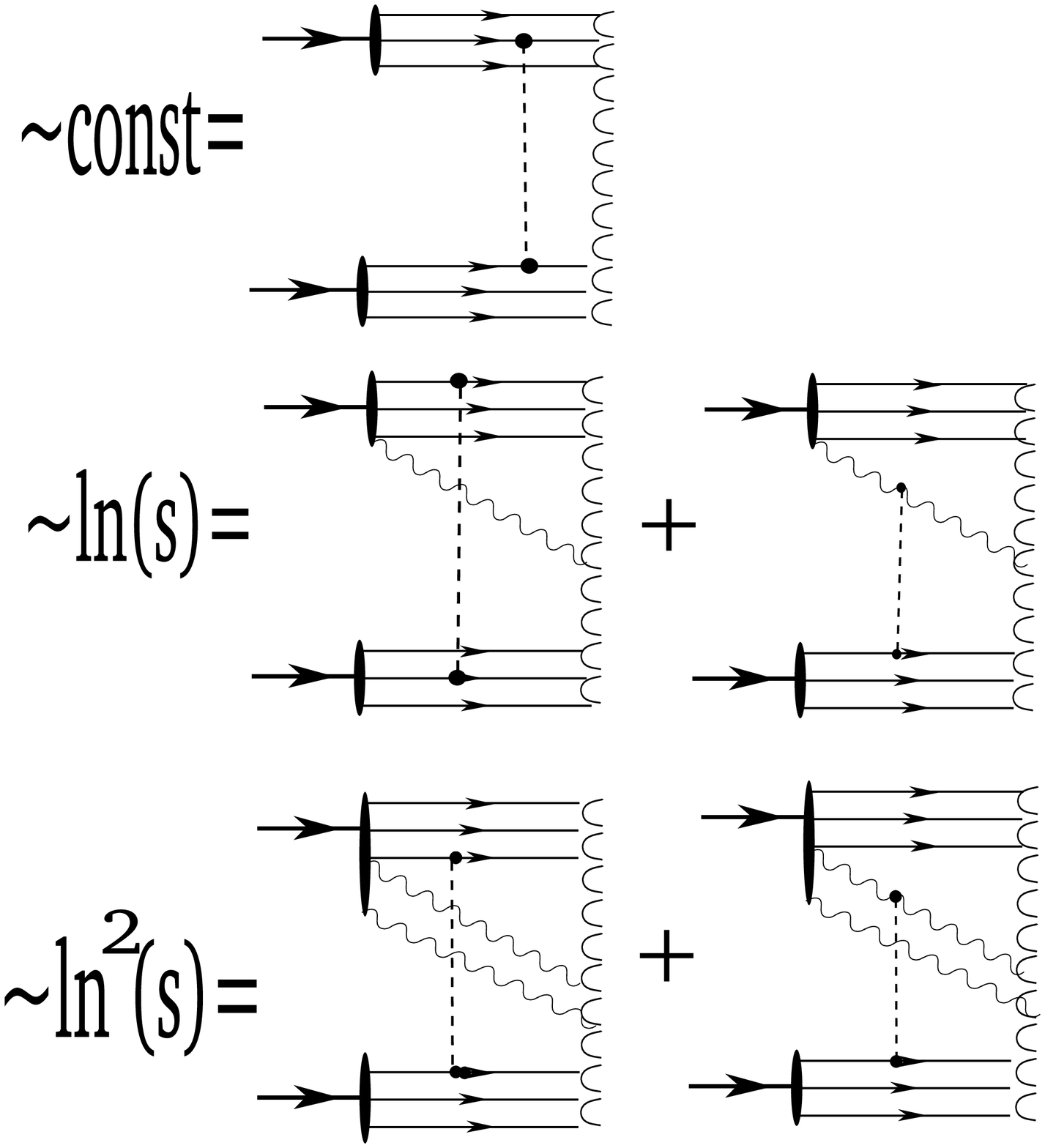}
\vline
\hspace{0.1in}
\includegraphics[width=2.5in]{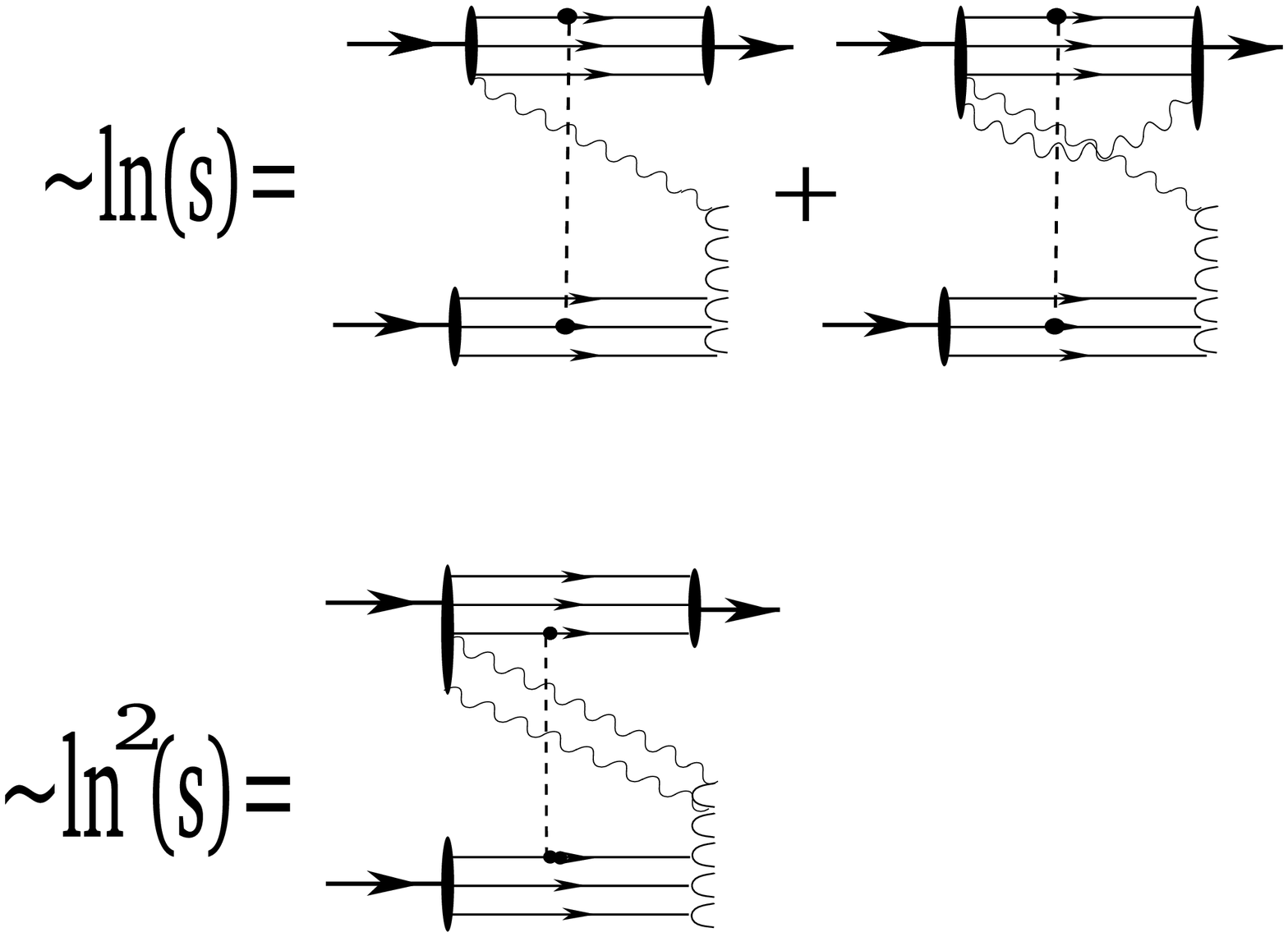}
\caption{Leading by powers of $ln(s)$ diagrams in low constituent model for total cross section (left part) and single diffraction cross section (right part).}
\label{fig:sn}
\end{figure}

This processes are schematically drawn on the left side of Fig.\ref{fig:sn}. 

Processes of the single difraction are naturally discribed in this model. To get colourless final state of the incoming hadron, part of the incoming bremsstralung must be colourless in the sum with exchanged coulomb gluon. So, the part of the longtitudual momenta of the incoming hadron  $x=\frac{M^2}{s}$, which is equal to the sum of $x_i$ exchaged bremsstralung gluons, will be transferred. At the final stage after hadronization we get the system composed of the leading hadron, lost part of the moments $x$, and rapidity separated hadrons with mass $M^2=s(1-x)$. This processes are schematically drawn on the right side of Fig.\ref{fig:sn}.

To estimate the rise of cross section of single diffraction with energy, we will compare leading terms of $ln(s)$ for single diffraction and total cross sectons. Leading diagrams for processes of single diffraction are the same as in the case of the total cross section. To describe processes of single diffraction we must fix colours of exchanged gluons to get colourless final states. This condition cancels constant term in sum of $ln(s)$ powers for single diffraction cross section. In the case of the single diffraction coefficient at $ln(s)$ will be suppressed by factor $\frac{1}{8}$ in comparision to expression for total cross section, because of $t$-channel exchange colourless condition. Coefficient at $ln^2(s)$ will be suppressed by factor $\frac{1}{256}$. Coefficients at different powers of  $ln(s)$ will be also supressed by kinematical factor $K$. It arises because the area of integration on longtitudual momentums of exchanged gluons is limited by the condition, that full exchanged momentum must be in the range $\frac{1.4 GeV^2}{s}..0.15$ by the common definition of $\sigma_{SD}$.  We assume, that these kinematical factors are the same for coefficients at $ln(s)$ and $ln^2(s)$. Therefore, if we take decomposition of cross sections on powers of  $ln(s)$ like
\begin{eqnarray}
\sigma_{tot}=A_1 + B_1 ln(s) + C_1 ln^2(s)\\
\sigma_{SD}=B_2 ln(s) + C_2 ln(s)^2
\end{eqnarray}
then the next conditions take place
\begin{eqnarray}
B_2=\frac{K}{8}B_1 \label{eqn:c1}\\
C_2=\frac{K}{256}C_1 \label{eqn:c2}
\end{eqnarray}
\begin{equation}
F^{color suppresion} \equiv \frac{B_2}{B_1}/\frac{C_2}{C_1}=32
\label{eq:fract}
\end{equation}
Last fraction we can call the colour suppression factor. From the CDF\cite{CDF} experimental data we have $B_1=1.147$,$C_1=0.1466$, and from total cross section data analysis we have $B_2=0.598$, $C_2=0.00214$. For the relation  interested to us (\ref{eq:fract}) we have
\begin{equation}
F^{colour suppresion}_{experimental} \equiv \left( \frac{B_2}{B_1}/\frac{C_2}{C_1} \right) _{experimental} \sim 36
\end{equation}
Precision of the last value is low, but we can state, that $F^{colour suppression}_{min} \sim 10$. Therefore term proportional $ln^2(s)$ is highly suppressed in comparsion with the total cross section, and this effect can be simply explained in our model by existence of color factor in (\ref{eqn:c1})-(\ref{eqn:c2}).

This model gives the same values of pomeron slope for elastic and single diffraction cross sections. This slope value is $\alpha^{\prime} \sim 0.16..0.26$ and consistent with CDF data. This fact is another confirmation of the low constituent model.

\section*{Acknowledgments}
We thank N.V.Prikhod`ko for useful discussions. This work was supported by RFBR Grant RFBR-03-02-16157a and Grant of Ministry for educations E02-3.1-282


\end{document}